\documentstyle[twoside,psfig,epsfig]{article}

\catcode`\@=11
\long\def\@makefntext#1{
\protect\noindent \hbox to 3.2pt {\hskip-.9pt  
$^{{\eightrm\@thefnmark}}$\hfil}#1\hfill}		

\def\thefootnote{\fnsymbol{footnote}}
\def\@makefnmark{\hbox to 0pt{$^{\@thefnmark}$\hss}}	
	
\def\ps@myheadings{\let\@mkboth\@gobbletwo
\def\@oddhead{\hbox{}
\rightmark\hfil\eightrm\thepage}   
\def\@oddfoot{}\def\@evenhead{\eightrm\thepage\hfil
\leftmark\hbox{}}\def\@evenfoot{}
\def\sectionmark##1{}\def\subsectionmark##1{}}

\oddsidemargin=\evensidemargin
\renewcommand{\thefootnote}{\fnsymbol{footnote}}

\newcounter{sectionc}\newcounter{subsectionc}\newcounter{subsubsectionc}
\renewcommand{\section}[1] {\vspace{12pt}\addtocounter{sectionc}{1} 
\setcounter{subsectionc}{0}\setcounter{subsubsectionc}{0}\noindent 
	{\tenbf\thesectionc. #1}\par\vspace{5pt}}
\renewcommand{\subsection}[1] {\vspace{12pt}\addtocounter{subsectionc}{1} 
	\setcounter{subsubsectionc}{0}\noindent 
	{\bf\thesectionc.\thesubsectionc. {\kern1pt \bfit #1}}\par\vspace{5pt}}
\renewcommand{\subsubsection}[1] {\vspace{12pt}\addtocounter{subsubsectionc}{1}
	\noindent{\tenrm\thesectionc.\thesubsectionc.\thesubsubsectionc.
	{\kern1pt \tenit #1}}\par\vspace{5pt}}
\newcommand{\nonumsection}[1] {\vspace{12pt}\noindent{\tenbf #1}
	\par\vspace{5pt}}

\newcounter{appendixc}
\newcounter{subappendixc}[appendixc]
\newcounter{subsubappendixc}[subappendixc]
\renewcommand{\thesubappendixc}{\Alph{appendixc}.\arabic{subappendixc}}
\renewcommand{\thesubsubappendixc}
	{\Alph{appendixc}.\arabic{subappendixc}.\arabic{subsubappendixc}}

\renewcommand{\appendix}[1] {\vspace{12pt}
        \refstepcounter{appendixc}
        \setcounter{figure}{0}
        \setcounter{table}{0}
        \setcounter{lemma}{0}
        \setcounter{theorem}{0}
        \setcounter{corollary}{0}
        \setcounter{definition}{0}
        \setcounter{equation}{0}
        \renewcommand{\thefigure}{\Alph{appendixc}.\arabic{figure}}
        \renewcommand{\thetable}{\Alph{appendixc}.\arabic{table}}
        \renewcommand{\theappendixc}{\Alph{appendixc}}
        \renewcommand{\thelemma}{\Alph{appendixc}.\arabic{lemma}}
        \renewcommand{\thetheorem}{\Alph{appendixc}.\arabic{theorem}}
        \renewcommand{\thedefinition}{\Alph{appendixc}.\arabic{definition}}
        \renewcommand{\thecorollary}{\Alph{appendixc}.\arabic{corollary}}
        \renewcommand{\theequation}{\Alph{appendixc}.\arabic{equation}}
        \noindent{\tenbf Appendix \theappendixc #1}\par\vspace{5pt}}
\newcommand{\subappendix}[1] {\vspace{12pt}
        \refstepcounter{subappendixc}
        \noindent{\bf Appendix \thesubappendixc. {\kern1pt \bfit #1}}
	\par\vspace{5pt}}
\newcommand{\subsubappendix}[1] {\vspace{12pt}
        \refstepcounter{subsubappendixc}
        \noindent{\rm Appendix \thesubsubappendixc. {\kern1pt \tenit #1}}
	\par\vspace{5pt}}

\newcommand{\textlineskip}{\baselineskip=13pt}
\newcommand{\smalllineskip}{\baselineskip=10pt}

\def\eightcirc{
\begin{picture}(0,0)
\put(4.4,1.8){\circle{6.5}}
\end{picture}}
\def\eightcopyright{\eightcirc\kern2.7pt\hbox{\eightrm c}}

\def\abstracts#1#2#3{{
	\centering{\begin{minipage}{4.5in}\baselineskip=10pt\footnotesize
	\parindent=0pt #1\par 
	\parindent=15pt #2\par
	\parindent=15pt #3
	\end{minipage}}\par}} 

\renewenvironment{thebibliography}[1]
	{\frenchspacing
	 \ninerm\baselineskip=11pt
	 \begin{list}{\arabic{enumi}.}
        {\usecounter{enumi}\setlength{\parsep}{0pt}     
	 \setlength{\leftmargin 12.7pt}{\rightmargin 0pt} 
         \setlength{\itemsep}{0pt} \settowidth
	{\labelwidth}{#1.}\sloppy}}{\end{list}}

\newcounter{itemlistc}
\newcounter{romanlistc}
\newcounter{alphlistc}
\newcounter{arabiclistc}

\newcommand{\fcaption}[1]{
        \refstepcounter{figure}
        \setbox\@tempboxa = \hbox{\footnotesize Fig.~\thefigure. #1}
        \ifdim \wd\@tempboxa > 5in
           {\begin{center}
        \parbox{5in}{\footnotesize\smalllineskip Fig.~\thefigure. #1}
            \end{center}}
        \else
             {\begin{center}
             {\footnotesize Fig.~\thefigure. #1}
              \end{center}}
        \fi}

\newcommand{\tcaption}[1]{
        \refstepcounter{table}
        \setbox\@tempboxa = \hbox{\footnotesize Table~\thetable. #1}
        \ifdim \wd\@tempboxa > 5in
           {\begin{center}
        \parbox{5in}{\footnotesize\smalllineskip Table~\thetable. #1}
            \end{center}}
        \else
             {\begin{center}
             {\footnotesize Table~\thetable. #1}
              \end{center}}
        \fi}

\def\@citex[#1]#2{\if@filesw\immediate\write\@auxout
	{\string\citation{#2}}\fi
\def\@citea{}\@cite{\@for\@citeb:=#2\do
	{\@citea\def\@citea{,}\@ifundefined
	{b@\@citeb}{{\bf ?}\@warning
	{Citation `\@citeb' on page \thepage \space undefined}}
	{\csname b@\@citeb\endcsname}}}{#1}}

\newif\if@cghi
\def\cite{\@cghitrue\@ifnextchar [{\@tempswatrue
	\@citex}{\@tempswafalse\@citex[]}}
\def\citelow{\@cghifalse\@ifnextchar [{\@tempswatrue
	\@citex}{\@tempswafalse\@citex[]}}
\def\@cite#1#2{{$\null^{#1}$\if@tempswa\typeout
	{IJCGA warning: optional citation argument 
	ignored: `#2'} \fi}}

\def\pmb#1{\setbox0=\hbox{#1}
	\kern-.025em\copy0\kern-\wd0
	\kern.05em\copy0\kern-\wd0
	\kern-.025em\raise.0433em\box0}

\def\fnt#1#2{\footnotetext{\kern-.3em
	{$^{\mbox{\scriptsize #1}}$}{#2}}}

\def\fpage#1{\begingroup
\voffset=.3in
\thispagestyle{empty}\begin{table}[b]\centerline{\footnotesize #1}
	\end{table}\endgroup}

\font\tenrm=cmr10
\font\tenit=cmti10 
\font\tenbf=cmbx10
\font\bfit=cmbxti10 at 10pt
\font\ninerm=cmr9

\font\eightrm=cmr8

\def\qed{\hbox{${\vcenter{\vbox{			
   \hrule height 0.4pt\hbox{\vrule width 0.4pt height 6pt
   \kern5pt\vrule width 0.4pt}\hrule height 0.4pt}}}$}}

\renewcommand{\thefootnote}{\fnsymbol{footnote}}	

\begin{document}
\newcommand{\qt}[1]{``{\em #1}''}
\newcommand{\lsim}{\raisebox{-0.13cm}{~\shortstack{$<$ \\[-0.07cm] $\sim$}}~}
\newcommand{\gsim}{\raisebox{-0.13cm}{~\shortstack{$>$ \\[-0.07cm] $\sim$}}~}

\newcommand{\dx}{\mbox{\rm d}}
\newcommand{\ra}{\rightarrow}
\newcommand{\ee}{e^+e^-}
\newcommand{\s}{\\ \vspace*{-3.5mm} }
\newcommand{\nn}{\noindent}
\newcommand{\non}{\nonumber}
\newcommand{\beqn}{\begin{eqnarray}}
\newcommand{\eeqn}{\end{eqnarray}}
\newcommand{\SM}{SM}
\newcommand{\SUSY}{SUSY}
\newcommand{\MSSM}{MSSM}
\newcommand{\SMG}{ {\rm SU(3) \times SU(2)\times U(1)}}
\newcommand{\tg}{\mbox{tan}}
\newcommand{\tb}{\mbox{tan}\beta}
\newcommand{\ctb}{\mbox{ctg}\!\beta}
\newcommand{\?}{ {\bf ???}}
\newcommand{\ctg}{\mbox{ctg}}
\newcommand{\MeV}{MeV}
\newcommand{\GeV}{\rm GeV}
\newcommand{\TeV}{TeV}
\def\inpb{\mbox{$\hbox{pb}^{-1}$}}
\def\Gcs{\hbox{GeV}/\mbox{$c^2$}}
\font\ninerm=cmr9


\begin{flushright}
PM/98--18\\
GDR--S--013
\end{flushright}
\vspace*{0.4cm}

\normalsize\textlineskip
\thispagestyle{empty}
\setcounter{page}{1}

\fpage{1}
\centerline{\bf IMPACT OF THE SUSY DECAYS ON THE SEARCH FOR}
\vspace*{0.035truein}
\centerline{\bf THE MSSM HIGGS BOSONS AT THE LHC} 
\vspace*{0.37truein}
\centerline{\footnotesize A. DJOUADI}
\vspace*{0.015truein}
\centerline{\footnotesize\it Laboratoire de Physique Math\'ematique et 
Th\'eorique, UMR--CNRS 5825}
\baselineskip=10pt
\centerline{\footnotesize\it Universit\'e Montpellier II, F--34095 
Montpellier Cedex 5, France.}
\vspace*{0.225truein}


\vspace*{0.21truein}
\abstracts{In the context of the Minimal Supersymmetric extension of the 
Standard Model, we discuss the impact of the decays of the neutral Higgs 
bosons into supersymmetric particles, charginos/neutralinos and sfermions. 
We show that these decay modes could be dominant, when they are kinematically 
accessible, thus strongly suppressing the branching ratios for the decay
channels which are used to detect the Higgs bosons at hadron colliders. 
These SUSY decay modes should therefore not be overlooked in the search for 
the Higgs particles at the LHC.}{}{}



\vspace*{1pt}\textlineskip	
\section{Introduction}  	
\vspace*{-0.5pt}
\noindent

\textheight=7.8truein
\setcounter{footnote}{0}
\renewcommand{\thefootnote}{\alph{footnote}}

The search for Higgs bosons\cite{HHG} is one of the main entries in the LHC 
agenda. While the search for the Standard Model (SM) Higgs bosons has been 
shown to be rather straightforward, provided that a high luminosity $\int {\cal
L} \sim 300$ fb$^{-1}$ is collected by the ATLAS and CMS 
collaborations\cite{LHCex}, 
the search for the Higgs particles of supersymmetric extensions 
(SUSY) of the SM, seems to be slightly more involved.\cite{LHCex,LHCth} 
The simplest version, the Minimal Supersymmetric Standard Model (MSSM), leads
to the existence of five physical states: two CP--even Higgs bosons $h$ and $H$,
a CP--odd Higgs boson $A$ and two charged Higgs particles $H^\pm$.\cite{HHG} 
In principle, many of the numerous decay modes and production processes of the 
MSSM Higgs bosons are needed to cover the full MSSM parameter space; see 
Ref.\cite{LHCcont} 

However, in many scenarii such as mSUGRA,\cite{LHCth} the MSSM Higgs sector is 
in the so called decoupling regime for most of the SUSY parameter space allowed
by present data constraints.\cite{data} The heavier $H,A$ and $H^\pm$ states 
are rather heavy and degenerate in mass, while the lightest $h$ boson reaches 
its maximal allowed mass value $ M_h \lsim $ 80--130 GeV\cite{mh} and has 
almost the same properties as the SM Higgs boson. In a large part of the MSSM 
parameter space, only the lightest $h$ boson can be produced at the LHC 
and, with some luck, the heavy $H,A$ particles. 

At the LHC, the most promising channel\cite{LHCex,LHCth} for detecting the
lightest $h$ boson is the rare decay into two photons, $h \rightarrow \gamma 
\gamma$, with the Higgs particle dominantly produced  via the top quark loop 
mediated gluon--gluon fusion mechanism\cite{gg} $gg  \rightarrow 
h$\footnote{ Two other channels can also be used to detect the $h$ particle 
in this mass range: the production in association with a $W$ boson\cite{R8} 
or with top quark pairs\cite{R9} leading to clean $\gamma \gamma + l^\pm$ 
events.}. In the 
decoupling regime, the two LHC collaborations expect to detect the narrow 
$\gamma \gamma$ peak in the entire Higgs mass range\footnote{Note, however, 
that the rates in the $gg$ mechanism can be much smaller than expected 
if $\tilde{t}$ squarks are relatively light and their coupling to the $h$ boson 
strongly enhanced.\cite{ggh} Some compensation however might come from the 
process $pp \ra \tilde{t} \tilde{t}h$ which has rather large rates in 
this case.\cite{hstop}}, 80 $\lsim M_h \lsim 130$ GeV, with an integrated 
luminosity $\int {\cal L} \sim 300$ fb$^{-1}$ corresponding to three years of 
LHC running.\cite{LHCex} 

The heavy CP--even $H$ and CP--odd $A$ bosons can be searched for at the LHC 
through their decays modes into $\tau^+ \tau^-$ pairs with the Higgs bosons 
produced in the $gg$ fusion mechanism or in association with $b\bar{b}$ pairs: 
$gg \ra H/A$ and $gg, q \bar{q} \ra b\bar{b}+ H/A$.\cite{gg,R9} This needs 
large values $\tb \gsim 5$ for masses $M_{H,A} \gsim 300$ GeV, to enhance 
the production cross sections and the $\tau^+ \tau^-$ decay branching ratios, 
which in the case where only standard decay modes are allowed reach the 
asymptotic value of BR$(h \to \tau^+ \tau^-) \sim 10\%$. The decays into muon 
pairs, $H/A \ra \mu^+ \mu^-$, give a 
rather clean signal and can be used despite of the very small branching ratios,
which asymptotically reach the level of $\sim 4.10^{-4}$.\footnote{For lower 
values $\tb \lsim 3$, most 
of which will 
be covered by the upgrade of LEP2\cite{LEP2} to $\sqrt{s}=200$ GeV, and not 
too large $M_{H,A}$ values, the decays $H/A \ra t\bar{t}$, $H \ra hh \ra 
b\bar{b} \gamma \gamma$ and $A \ra Zh \ra l^+l^- b \bar{b}$ can also be used; 
see Ref.\cite{LHCcont}.}

However, in these analyses, it is always assumed that the heavy $H/A$ bosons 
decay only into standard particles, and that the SUSY decay modes are shut. 
But for such large values of $M_{H,A}$, at least 
the decays into the lightest neutralinos and charginos, and possibly into 
to light $\tilde{t}$ and $\tilde{b}$, can be kinematically allowed. These 
modes could have large decays widths, and thus could suppress the $H/A \ra
\tau^+ \tau^-$ branching ratios drastically. For the lightest $h$ boson, 
because of its small mass, only a little room is left for decays into SUSY 
particles by present experimental data.\cite{data} However, the 
possibility of $h$ decays into neutralinos  is not yet 
completely ruled out,
especially if one relaxes the gaugino mass unification; decays into sneutrinos
are also still possible. When these
invisible decays occur, they can be dominant, hence reducing the probability 
of the $h \ra \gamma \gamma$ decay to occur. These SUSY decays should therefore
not be overlooked as they might jeopardize the detection of the Higgs 
particles at the LHC. 

These SUSY decays of the Higgs bosons are discussed and updated in this note. 
All branching ratios are obtained with the help of an adapted version of the 
program HDECAY.\cite{R7} Previous analyses of SUSY Higgs boson decays 
can be found in Ref.\cite{R10} 

\section{Invisible decays of the h boson} 

Despite the lower bound of $91\,$GeV on the mass of the lightest chargino 
$\chi_1^+$ and the constraints from $\chi_0^1 \chi_0^2$ searches at 
LEP2\cite{data}, the decay of the lightest $h$ boson into a pair of
lightest neutralinos is still kinematically possible. Even in the constrained
MSSM with a common gaugino mass at the GUT scale, leading to the 
well--know relation between the wino and bino masses $M_1=\frac{5}{3} 
{\rm tg}^2\theta_W M_2\sim \frac{1}{2}M_2$, the lower bound on the LSP 
mass is only $m_{\chi_0^1} \gsim 30$ GeV.\cite{data} Since the upper 
bound on the lightest $h$ boson in the MSSM is $M_h \sim 130$ GeV,\cite{mh} 
there 
is still room for the invisible decay $h \ra \chi_0^1 \chi_0^1$ to occur. 

In the decoupling regime, the $hb\bar{b}$ coupling is SM--like and can be
much smaller than the $h\chi_1^0 \chi_1^0$ coupling; the decay of $h$ into the 
lightest neutralinos can be then dominant, resulting in a much smaller BR($h\ra 
\gamma \gamma$) than in the SM. Far from the decoupling limit, the coupling 
$g_{hbb} \sim \tan \beta$ is strongly enhanced for $\tb \gsim 3$, while the 
$h$ boson couplings to $W$ bosons and top quarks [which provide the main 
contributions to the $h \gamma \gamma$ loop vertex] are suppressed. This again 
will result in a strong suppression of BR($h \ra \gamma \gamma$).    

The partial width for the decay $h \ra \chi_0^1 \chi_0^1$ is given by
\beqn 
\Gamma (h \ra \chi_0^1 \chi_0^1)= \frac{G_F M_W^2 M_h}{2 \sqrt{2} \pi}
\ g^2_{h \chi_0^1 \chi_0^1} \ \beta^3_{\chi} 
\eeqn 
with $\beta_\chi^2=  1 - 4m^2_{\chi_1^0}/M_h^2$ and the normalized coupling 
$g_{h \chi_0^1 \chi_0^1}$ given by
\beqn
g_{h \chi_1^0 \chi_1^0} = (Z_{12}-\tan\theta_W Z_{11})(\sin \beta Z_{14} 
-\cos \beta Z_{13}) \label{gchi}
\eeqn 
with $Z$ is the matrix diagonalizing the neutralino mass 
matrix.\cite{HaberKane} The decay
is important only for moderate values of $M_2$ and $\mu$ [with a preference 
for $\mu>0$] since the $h$ boson prefers to couple to neutralinos which are a 
mixture of gauginos and higgsinos. In this range, the decay $h \ra \chi_0^1 
\chi_0^1$ is dominant if $M_h$ is above the $2m_{\chi_0^1}$ threshold; close 
to this value, the width is strongly suppressed by the $\beta^3_\chi$ factor. 

As an illustration of this possibility, we show in Fig.~1 the fraction
BR$(h \ra \gamma \gamma)$ as a function of $\mu$ for two values of $\tb=
2,30$. We choose $M_A = m_{\tilde{q}}= 1 $ TeV and the ``maximal mixing" 
scenario $A_t=\sqrt{6} m_{\tilde{q}}$ to maximize the $h$ boson mass  
[this gives $M_h \simeq 126$ GeV for $\tb=30$ and $ M_h \simeq 106$ GeV for 
$\tb=2$; the variation with $\mu$ is almost negligible]. In the $\tan \beta=30$
and $M_2=140$ GeV case, for  $|\mu| \gsim  200$ GeV the channel $h\ra \chi_1^0
\chi_1^0$ is kinematically closed and BR($h \to \gamma \gamma$) is SM--like, 
$\simeq 
2.3 \times 10^{- 3}$. In the range $110 \lsim |\mu| \lsim 200$ GeV, the LSP 
is lighter that $M_h/2$ while the chargino is still heavier than 91 GeV, 
the decay $h \ra \chi_1^0 \chi_1^0$ is thus allowed to occur and suppresses 
BR($h \to 
\gamma \gamma$). The suppression is stronger with decreasing $|\mu|$ since 
the phase--space becomes more favorable, and also the LSP tends to be an equal 
mixture of higgsino and gaugino. The maximum drop of BR($h \ra \gamma \gamma$) 
is a factor of three and two for $\mu>0$ and $\mu<0$ respectively. For values 
$|\mu| \lsim  110$ GeV, $m_{\chi_1^\pm}$ exceeds its experimentally allowed
lower bound. 

\begin{figure}[htb]
\vspace*{-.5cm}
\centerline{\mbox{\psfig{figure=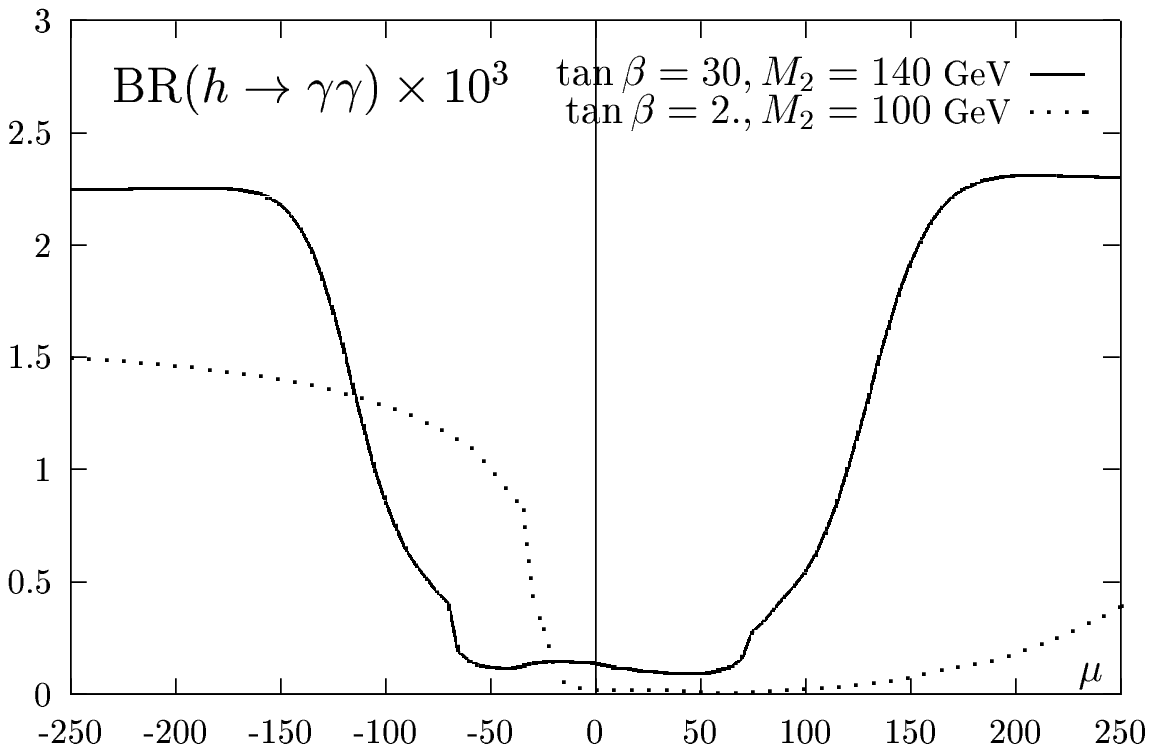,width=12cm}}}
\vspace*{-10.4cm}
\fcaption{Branching ratios in units of $10^{-3}$ for the decays $h \ra 
\gamma \gamma$ as a function of $\mu$ for $\tb=2(30)$ and $M_2=100(140)$ GeV.} 
\end{figure}

In the case $\tb=2$ and  $M_2=100$ GeV, the only experimentally allowed
region is $|\mu| \gsim 110$ GeV with $\mu<0$, since elsewhere the chargino is 
heavier than 91 GeV. In this $|\mu|$ range, the decay $h \ra \chi_1^0 \chi_1^0$
is kinematically allowed, but the branching ratio is very small, less than 
0.5\%. This is due to the fact that in this area $\chi_0^1$ is a pure bino 
state and its couplings to the $h$ boson are strongly suppressed. What makes
the $h \ra \gamma \gamma$ branching ratio drop by almost a factor two
compared to the previous case is first, the smaller value of $M_h$
[the decay width grows with the third power of the Higgs mass] and then 
because of the contribution of the chargino loops to the $h \ra \gamma 
\gamma$ decay, which interfere destructively for $\mu<0$ with the dominant 
contribution due $W$ boson loops [the reduction is nevertheless very mild, at 
most 15\% in this case]. 

If the constraint on the unification of the gaugino masses at the GUT scale
is relaxed, there is practically no lower bound on the LSP mass. Indeed,
for relatively large $\mu$ values, the lightest chargino $\chi_1^+$ and the 
next--to--lightest neutralino $\chi_2^0$ are wino--like with a mass $\sim M_2$ 
while the lightest neutralino is bino--like with a mass $\sim M_1$; since 
$M_1$ is a free parameter, it can be as small as possible leading to a 
possibly very light LSP. The decay $h \ra \chi_0^1 \chi_0^1$ will then have 
more room to occur. 

\begin{figure}[htb]
\vspace*{-.5cm}
\centerline{\mbox{\psfig{figure=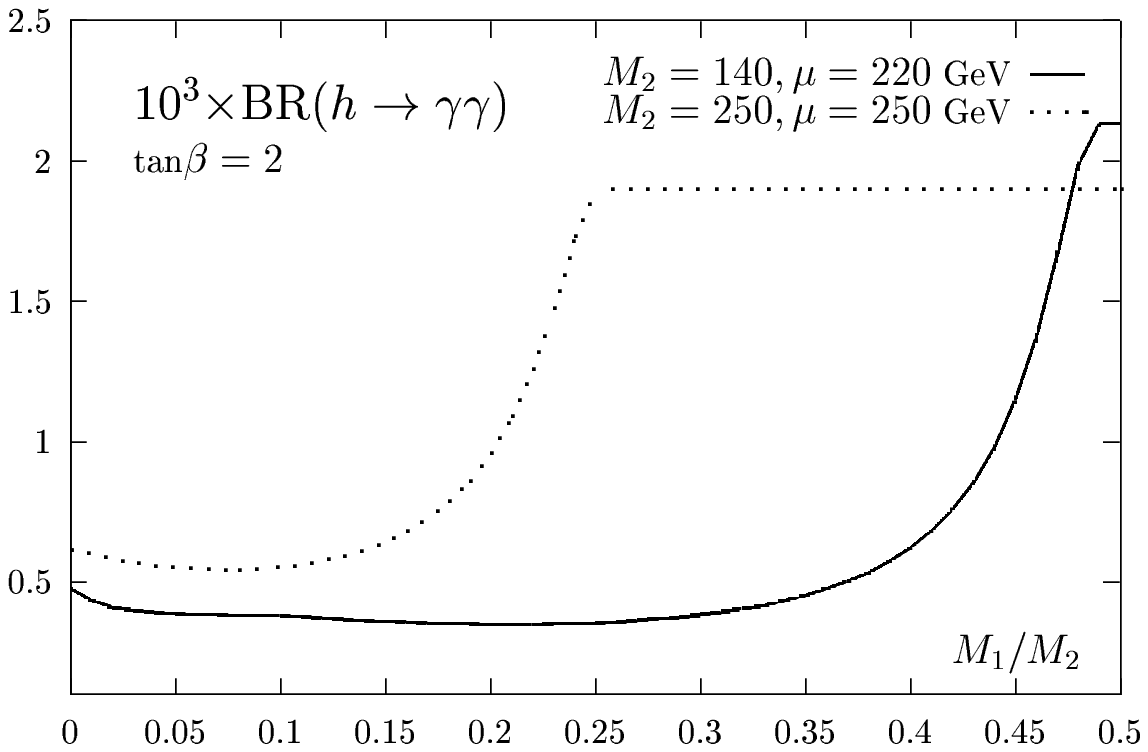,width=12cm}}}
\vspace*{-10.4cm}
\fcaption{Branching ratios for the decays $h \ra \gamma \gamma$ in units
of $10^{-3}$ as a function
of $M_1/M_2$ for $\tb=2$ and two sets of $M_2,\mu$ values.} 
\end{figure}

The branching ratio for the decay $h \ra \chi_1^0 \chi_1^0$ can be rather 
large thus suppressing the $\gamma \gamma$ branching ratio. This is exemplified 
in Fig.~2, where BR($h\ra \gamma \gamma$) is shown as a function of the 
ratio $M_1/M_2$ for $\tb=2$ and for two sets of $M_2$ and $\mu$ values;
$M_2=140$ GeV, $\mu=220$ GeV leading  to $m_{\chi_1^+}\simeq 96$ GeV, and 
$M_2=\mu=250$ GeV leading to $m_{\chi_1^+}\simeq 175$ GeV; the remaining 
inputs are as in the previous figure. 
In the first scenario, when the LSP is very light BR$(h \ra 
\gamma \gamma$) drops to the level of $5.10^{-4}$, a strong reduction compared
to the expected rate $\sim 2.10^{-3}$. With increasing $M_1/M_2$ and hence
with increasing LSP mass, it stays almost constant until the $2m_{\chi_1^0}$ 
threshold is reached for $M_1 \sim M_2/2$ and the rate recovers its
standard value. In the second scenario,  
BR($h\ra \gamma \gamma$) starts at the same level 
as previously, but increases more
rapidly and reaches approximately the standard value for $M_1 \sim M_2/4$ 
which corresponds to the kinematical limit for the decay $h \ra \chi_1^0 
\chi_1^0$. When the LSP decay is shut, the difference between the $\gamma
\gamma$ branching ratios in the two scenarios is due to a constructively 
interfering chargino loop contribution [the sign of the $\chi_1^\pm$ 
contribution goes with the sign of $\mu$] in the case where $m_{\chi_1^+}
\simeq 96$ GeV and which enhances the $\gamma \gamma$ decay width by 20\% 
or so. 
This picture is expected not to be altered significantly for larger values of 
$\tb$ if the $h$ boson is in the decoupling regime as discussed previously 
[in fact for large $\tb$ values and for some moderate values of the parameters 
$M_2$ and $\mu$, even the decays into the lightest and the next to lightest 
neutralinos is possible]. 

Another kinematically still possible SUSY mode for the lightest $h$ boson is 
the decay into sneutrinos. Indeed, the experimental lower bound on the $\tilde{
\nu}$ masses is still rather low, $m_{\tilde{\nu}} \gsim 45$ GeV,\cite{data} 
leaving some room for the decay $h \ra \tilde{\nu} \tilde{\nu}$ to occur. 
However, because of SU(2)$_{\rm L}$ invariance, the sneutrino and the 
left--handed charged slepton masses are related and one should avoid being 
into conflict with the stronger experimental bound $m_{\tilde{l}_L} \gsim 70$ 
GeV. However, even in this case one can obtain a rather light sneutrino since 
a splitting between the $\tilde{\nu}$ and $\tilde{l}_L$ masses can be generated
by the D--terms. Indeed, denoting the common scalar mass by $\tilde{m}$, one 
has:
\beqn
m_{\tilde{\nu}}^2 \simeq \tilde{m}^2 +0.50 M_Z^2 \cos2\beta \ \ , \ \ 
m_{\tilde{l}_L}^2 \simeq \tilde{m}^2 -0.27 M_Z^2 \cos2\beta
\eeqn
For small values of $\tilde{m}$, the slepton masses are governed by the 
D--terms, and for large values of $\tb$, $\cos 2\beta \ra -1$ and the 
D--terms become maximal. Since they tend to increase $m_{\tilde{l}_L}$
and decrease $m_{\tilde{\nu}}$, relatively low masses for sneutrinos can be 
kept while still having rather heavy left--handed\footnote{The D--terms
for right--handed charged sleptons are approximately the same as for
the left--handed ones and tend also to decrease the mass. However,
in GUT scenarii such as mSUGRA, the $\tilde{l}_R$ tends to be lighter 
than the sneutrinos for reasonable values of the gaugino mass $m_{1/2}$. 
In this case, the decay $h \ra \tilde{\nu} 
\tilde{\nu}$ is forbidden because of the experimental bound $m_{\tilde{l}_R} 
\gsim 70$ GeV.}~~sleptons [note however, that $\tilde{\nu}$ should not be
lighter than the lightest neutralino which is expected to be the LSP]. 

In the decoupling limit, the $h$ boson coupling to sneutrinos is also 
proportional to $\cos 2\beta$, and for large 
$\tb$ values it becomes maximal. And since it is a  ``gauge" coupling, 
it is much larger than the $hb\bar{b}$ Yukawa coupling, and the decay $h \ra 
\tilde{\nu} \tilde{\nu}$ is always largely dominating once it is kinematically 
allowed. The partial width for the decay, summing over the three 
sneutrinos, is given by
\beqn 
\Gamma (h \ra \tilde{\nu} \tilde{\nu})= \frac{3 G_F M_Z^4}{8 \sqrt{2} \pi M_h}
\beta_{\tilde{\nu}} \ \ , \ \ \beta_{\tilde{\nu}} = \left[ 1 - \frac{4 
m^2_{\tilde{\nu}}}{M_h^2} \right]^{1/2} 
\eeqn 
Modulo the velocity factor $\beta_{\tilde{\nu}}$, 
the partial width is larger than the otherwise dominant
$b\bar{b}$ decay width by a huge factor: $M_Z^4/(2 m_b^2 M_h^2) \sim  
230$ for $M_h=130$ GeV. Thus, if the  $h \ra \tilde{\nu} \tilde{\nu}$ decay
mode is allowed, all the branching ratios for the other decay channels
including the $h \ra \gamma \gamma$ mode, will be suppressed by two
orders of magnitude. Since
the sneutrinos will decay invisibly in this mass range [$m_{\tilde{\nu}} < 
m_{\chi_1^\pm}$ and the only possible channel  is the invisible mode 
$\tilde{\nu} \ra 
\nu \chi_1^0$],  the $h$ boson would be then also very difficult to detect at 
the LHC\footnote{At $e^+ e^-$ colliders missing mass techniques allow for an 
easy detection in the process $e^+ e^- \ra hZ$.}. 

\section{H/A decays into SUSY particles} 

If the CP--even and the CP--odd Higgs bosons $H$ and $A$ are heavy, $M_{H,A}
\gsim 300$ GeV, at least the decays into the lightest neutralinos and possibly 
charginos should be kinematically allowed. For moderate $\tb$ values, the 
couplings to $b\bar{b}$ and $\tau^+ \tau^-$ pairs [which together with $t
\bar{t}$ states account for the total width in the absence of SUSY modes] 
are not strongly enhanced, these decays might be dominant and suppress 
drastically the branching ratios for the $H/A \ra \tau^+ \tau^-$ signals. 
The partial widths for the decays of the particle $\Phi =H,A$ into 
$\chi_i \chi_j$ states are given by\cite{R10}
\begin{small}
\begin{eqnarray}
\Gamma =  \frac{G_F M_W^2 M_\Phi \lambda_{ij}^{1/2} }
{2 \sqrt{2} \pi (1+\delta_{ij}) } 
\left[ (F_{ij \Phi}^2 + F_{ji \Phi}^2) \left(1- \frac{ m_{\chi_i}^2}
{M_{\Phi}^2} - \frac{ m_{\chi_j}^2}{M_{\Phi}^2} \right) 
-4\eta_\Phi \epsilon_i \epsilon_j F_{ij\Phi} F_{ji\Phi} \frac{ 
m_{\chi_i} m_{\chi_j}} {M_{\Phi}^2} \right]
\end{eqnarray}
\end{small}
where $\eta_{H}=+1$, $\eta_A=-1$ and $\delta_{ij}=0$ unless the
final state consists of two identical (Majorana) neutralinos in which
case $\delta_{ii}=1$; $\epsilon_i =\pm 1$ stands for the sign of the 
$i$'th eigenvalue of the neutralino mass matrix\cite{HaberKane}
while  $\epsilon_i=1$ for charginos; 
$\lambda_{ij} = (1-m_{\chi_i}^2/M_\Phi^2- m_{\chi_j}^2
/M_\Phi^2)^2-4m_{\chi_i} ^2 m_{\chi_j}^2/M_\Phi^4$. The coefficients 
$F_{ij\Phi}$ are related to the elements of the matrices\cite{HaberKane} 
$U,V$ and $Z$ for charginos and neutralinos; in the decoupling limit they 
are the same for $H$ and $A$ and read
\begin{eqnarray}
\Phi \ra \chi_i^0 \chi_j^0 \ \ &:& F_{ij\Phi}= \frac{1}{2}
\left( Z_{j2}- \tan\theta_W Z_{j1} \right) \left( \sin \beta Z_{i3} -  
\cos \beta Z_{i4} \right) \ + \ i \leftrightarrow j \non \\
\Phi \ra \chi_i^+ \chi_j^- \ &:& F_{ij\Phi }= \frac{1}{\sqrt{2}} \left[
\sin \beta V_{i1}U_{j2} + \cos \beta  V_{i2}U_{j1} \right] 
\end{eqnarray}

\begin{figure}[htb]
\vspace*{-.5cm}
\centerline{\mbox{\psfig{figure=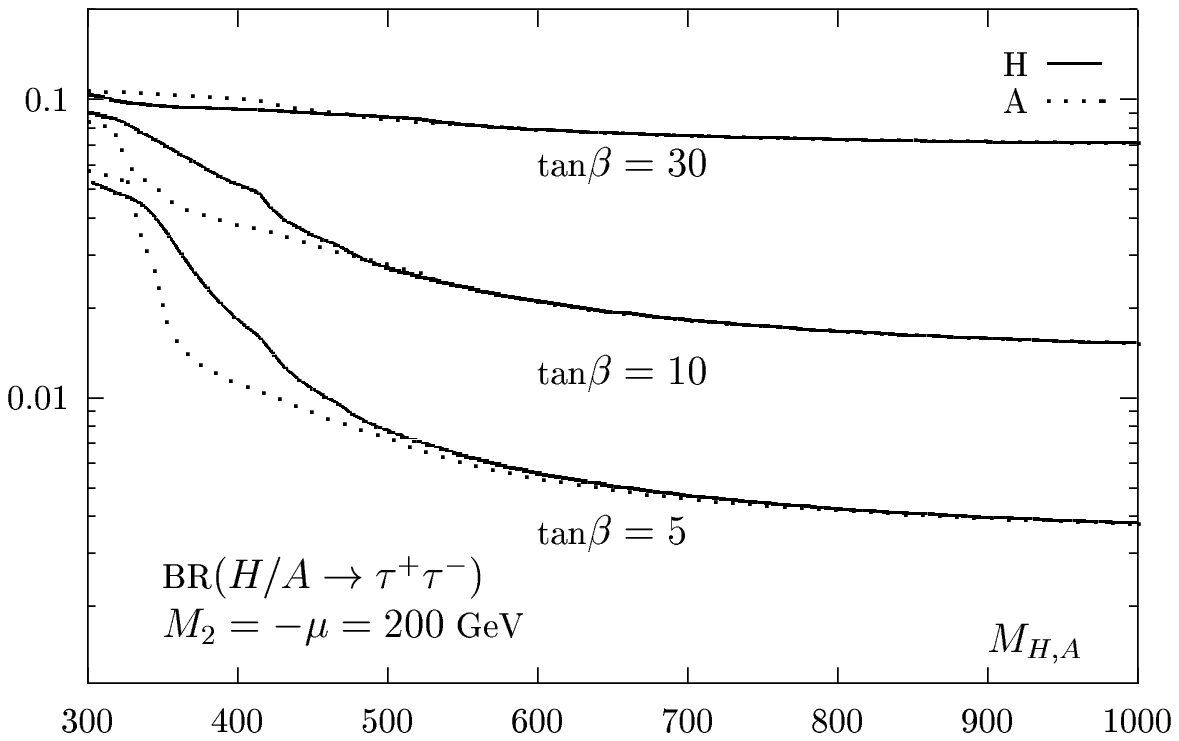,width=12cm}}}
\vspace*{-10.4cm}
\fcaption{Branching 
ratios for the decays $H/A \ra \tau^+ \tau^-$ as a function
of $M_{H,A}$ for $\tb=5,10$ and 30 and for the values $M_2=-\mu=200$ GeV.} 
\end{figure}

In Fig.~3, the branching fractions  BR$(H/A \ra \tau^+\tau^-)$ are plotted as 
a function of the $H/A$ masses for three values $\tb=5,10$ and 30. The choice 
$M_2=-\mu=200$ GeV has been made leading to $m_{\chi_1^0} \sim 90$ GeV and 
$m_{\chi_1^+} \sim 160$ GeV [with a small variation with $\tb$]. 
The branching ratios for $H$ and $A$ decays are almost the same except for 
small values of $\tb$ and relatively small Higgs masses: in this case, the 
decoupling limit is not yet reached and additional [and different] decay 
modes occur for the $H$ and $A$ bosons as discussed previously. For $\tb=5$
the $H/A$ couplings to down--type fermions are not very strongly enhanced
and the decays into charginos and neutralinos have large branching ratios: 
they decrease BR$(H/A \ra \tau^+ \tau^-)$ from the standard $\sim 10\%$ 
value for small Higgs masses [where only a few SUSY channels are open and 
some are suppressed by phase space] to less than 0.4\% for very heavy Higgs 
boson masses $M_{H,A} \sim 1$ TeV [here most of the neutralino/chargino
channels are open and they are not suppressed by phase space], thus a 
reduction by more than a factor of 20 compared to the branching ratio without 
the SUSY decays. For $\tb=10$, the couplings to $b$--quarks and $\tau$--leptons
are more enhanced and BR$(H/A \ra \tau^+\tau^-)$ are larger by slightly more 
than a factor of two compared to the previous case. For even larger values of 
$\tb$, $\tb=30$, the decays into charginos and neutralinos are not dominating 
anymore, and the branching ratios for the $H/A$ decays into tau pairs are 
suppressed only slightly, less than a factor of two. 

In the preceding discussion, the decays of $H$ and $A$ into 
sfermions were assumed to be shut. However, at least the decays into the 
lightest stops can be kinematically allowed, and strongly 
enhanced\footnote{This 
might also be the case of the sbottoms for large values of $\tb$ and the 
parameters $\mu$ and $A$; however this will not be discussed here and we will 
assume that the mixing is zero in this sector.}. Indeed, the current stop
eigenstates, $\tilde{t}_L$ and $\tilde{t}_R$, mix to give the mass eigenstates 
$\tilde{t}_1$ and $\tilde{t}_2$; the mixing angle $\theta_{\tilde{t}}$ is 
proportional to $A_t - \mu/\tan \beta$, and can be very
large, leading to a $\tilde{t}_1$ much  lighter than the
$t$--quark and all other scalar quarks. In addition the couplings of the top 
squarks to the Higgs boson $H$ in the decoupling limit read 
\begin{eqnarray}
g_{H \tilde{t}_1 \tilde{t}_1 } = \sin 2\beta \left[ 
\frac{1}{2} \cos^2 \theta_{\tilde{t}} - \frac{2}{3} s^2_W \cos 2
\theta_{\tilde{t}} \right] - \frac{m_t^2}{M_Z^2} \frac{1}{\tb}
+ \frac{1}{2} \sin 2\theta_{\tilde{t}} \frac{m_t}{M_Z^2} (\frac{A_t}{\tb} 
+\mu)
\end{eqnarray}
For large values of $A_t$ or $\mu$, which incidentally make  $\theta_{
\tilde{t}}$ maximal, $|\sin 2 \theta_{\tilde{t}}| \simeq 1$, the 
last components can strongly  enhance the $g_{H\tilde{t} \tilde{t}}$ couplings 
and make them larger than the top quark coupling of the $H$ boson, $g_{Htt} 
\propto m_t/M_Z$. The pseudoscalar $A$ couples only to $\tilde{t}_1 \tilde{t}_2
$ pairs because of CP--invariance, the coupling is given by:
\begin{eqnarray}
g_{A \tilde{t}_1 \tilde{t}_2 } &=&  \frac{1}{2} \frac{m_t}{M_Z^2} 
(A_t/\tb -\mu) 
\end{eqnarray}
In the maximal mixing case, $|\sin2\theta_{\tilde{t}}| \simeq 1$, this is also 
the main component of the $H$ boson coupling to $\tilde{t}_1 \tilde{t}_2$ 
pairs except that the sign of $\mu$ is reversed. 

The partial decay widths of the $H,A$ bosons into top squarks are given by
\begin{eqnarray}
\Gamma (\Phi \ra \tilde{t}_i \tilde{t}_j ) = \frac{3 G_F }{2 \sqrt{2} 
\pi M_{\Phi} } \, \lambda^{1/2}_{ \tilde{f}_i \tilde{f}_j} \, 
g_{\Phi \tilde{t}_i \tilde{t}_j}^2
\end{eqnarray}
For the $H$ boson, the partial width, up to 
mixing angle factors is proportional  to $G_F m_t^4/$ $ (M_H \tan^2\beta)$ 
or/and $G_F m_t^2 (\mu -A_t/\tb)^2 /M_H$; for small $\tb$ values and  not too 
large $M_H$ and for intermediate $\tb$ values and for large $\mu$ and $A_t$, 
the width for the decays $H \ra \tilde{t} \tilde{t}$ can be very large and 
can compete with, and even dominate over, the other [standard and SUSY]
decay channels. The branching ratios for the $H$ decays into $\tau$ pairs 
would be then further suppressed. 

\begin{figure}[htbp]
\vspace*{-.5cm}
\centerline{\mbox{\psfig{figure=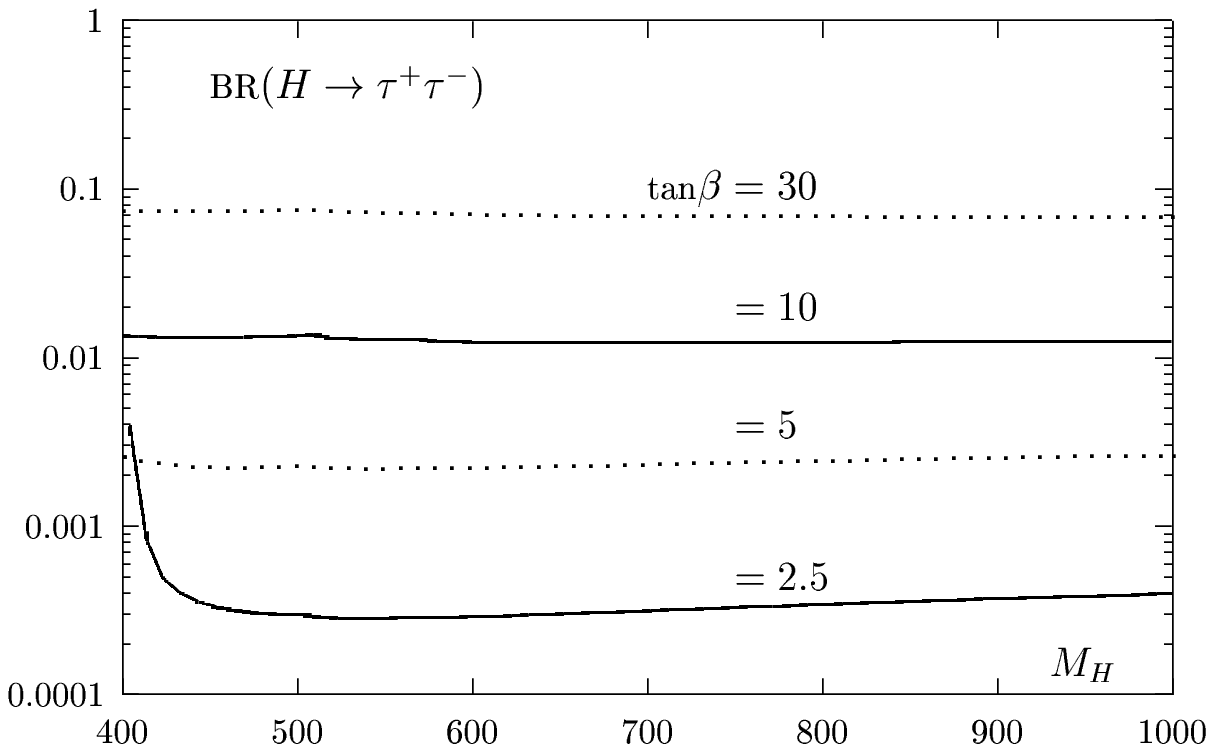,width=12cm}}} 
\vspace*{-10.4cm}
\fcaption{Branching fractions for $H \ra \tau^+ \tau^-$ as a function of 
$M_{H}$ for $\tb=2.5,5,10,30$, $m_{\tilde{t}_1} \simeq 200$ GeV, 
$M_2=\mu=m_{\tilde{f}}/2=250$ GeV and $A_t=1.5$ TeV.}
\end{figure}

This is illustrated in Fig.~4, where BR$(H\ra \tau^+ \tau^-$)
is shown as a function of $M_H$ for $\tb=2.5,5,10,30$ and  $m_{\tilde{t}_1} 
\simeq 200$ GeV [for $\tb=2.5$ this is achieved by setting $m_{\tilde{f}_L}
=m_{\tilde{f}_R} = 500$ GeV and $A_t =1.5$ TeV]; $M_2=\mu=250$ GeV. 
BR$(H \ra \tilde{t}_1 \tilde{t}_1$) decreases with increasing $\tb$ 
values and increasing $M_H$, but it is still at the level of $\sim 50\%$ 
for $\tb=5$ and $M_H=1$ TeV. For $\tb=30$, the channel $H \ra \tilde{t}_1 
\tilde{t}_2$ opens up for $M_H \sim 900$ GeV; however for this 
large $\tb$ value, the branching ratio barely exceeds the level of $20\%$ in 
contrast to lower $\tb$ values where it can reach almost unity for small $M_H$.
For larger $M_H$, the decays into charginos and neutralinos become more 
important and will dominate; so BR$(H \ra \tau^+ \tau^-)$ is reduced anyway. 

For the $A$ boson the only important decay into sfermions is $A\ra \tilde{t}_1
\tilde{t}_2$ [and maybe $\tilde{b}_1 \tilde{b}_2$ for $\tb \gg 1$]. Thus 
both stops must be light for the decay to be allowed by kinematics. This 
happens only in a small area of the parameter space, unless all squarks are 
relatively light. For instance, in the scenario above, $m_{\tilde{t}_2} 
\sim 700$ GeV and the decays $A,H \ra \tilde{t}_1 \tilde{t}_2$ occur only 
for masses close to 1 TeV. 
Note that the decay widths of the $H$ bosons into the light fermion
partners are proportional to $ G_F M_W^4 \sin^2 2\beta /M_{H}$ for $M_H \gg 
m_{\tilde{f}}$. They are thus suppressed by the heavy $H$ mass and cannot 
compete with the decays into fermions [$t, b, \tau$ and possibly $\chi$ states]
for which the widths grow as $M_H$. The pseudoscalar $A$ boson cannot decay 
into the partners of light fermions, if the fermion mass is neglected. 

\section{Conclusions} 

I have discussed the SUSY decay modes of the neutral Higgs bosons in the 
MSSM. Decays of the $h$ boson into invisible neutralinos [and also sneutrinos]
are still possible, especially if the gaugino mass unification constraint at 
the GUT scale is relaxed, and might be dominant when they occur, hence reducing 
the $h \ra \gamma \gamma$ branching ratio significantly. Decays of the $H$ and 
$A$ bosons into chargino and neutralino pairs, and decays of the $H$ boson 
into stops, are also important in large areas of the MSSM parameter space, 
and can suppress strongly the branching ratios for the $\tau^+ \tau^-$ 
discovery mode. These decay modes should not be overlooked, as they might 
jeopardize the search of the MSSM Higgs bosons at the LHC. 

\nonumsection{Acknowledgements}
\noindent
The work presented here has been initiated 
after stimulating discussions during the GDR--Supersym\'etrie workshop.

\nonumsection{References}

\end{document}